\documentclass[sigconf,authorversion,nonacm]{acmart}
\usepackage{tikz}
\def\checkmark{\tikz\fill[scale=0.4](0,.35) -- (.25,0) -- (1,.7) -- (.25,.15) -- cycle;}

\newcommand{\tikzxmark}{%
\tikz[scale=0.23] {
    \draw[line width=0.7,line cap=round] (0,0) to [bend left=6] (1,1);
    \draw[line width=0.7,line cap=round] (0.2,0.95) to [bend right=3] (0.8,0.05);
}}

\AtBeginDocument{%
  \providecommand\BibTeX{{%
    \normalfont B\kern-0.5em{\scshape i\kern-0.25em b}\kern-0.8em\TeX}}}

\newcommand*{\thead}[1]{%
\multicolumn{1}{|l|}{\begin{tabular}{@{}c@{}}#1\end{tabular}}}
\setcopyright{none}
\begin{document}

\title{Oracle-based Test Adequacy Metrics: A Survey}


\settopmatter{authorsperrow=3}

\author{Soneya Binta Hossain}
\email{sh7hv@virginia.edu}
\affiliation{%
  \institution{University of Virginia}
  \city{Charlottesville}
  \state{Virginia}
  \country{USA}
}
\author{Matthew B. Dwyer}
\email{matthewbdwyer@virginia.edu}
\affiliation{%
  \institution{University of Virginia}
  \city{Charlottesville}
  \state{Virginia}
  \country{USA}
}

\renewcommand{\shortauthors}{Hossain, et al.}

\begin{abstract}

Code coverage is a popular and widespread test adequacy metric
that measures the percentage of program codes executed by a test
suite. Despite its popularity, code coverage has several limitations.
One of the major limitations is that it does not provide any insights
into the quality or quantity of test oracles, a core component of
testing. Due to this limitation, several studies have suggested that coverage is a poor test adequacy metric; therefore, it should not be used as an indicator of a test suite's fault detection effectiveness. To address this limitation, researchers have proposed extensions to traditional structural code
coverage to explicitly consider the quality of test oracles. We refer to these extensions as \textit{oracle-based code coverage}. This survey paper studies oracle-based coverage techniques published since their inception in 2007.  We discuss each metric's definition, methodology, experimental studies, and research findings. Even though oracle-based coverage metrics are proven to be more effective than traditional coverage in detecting faults, they have received
little attention in the software engineering community. We present all existing oracle-based adequacy metrics in this paper and compare the critical features against each other. We observe that different oracle-based adequacy metrics operate on different coverage domains and use diverse underlying analysis techniques, enabling a software tester to choose the appropriate metric based on the testing requirements. Our paper provides valuable information regarding the limitations of oracle-based methods, addressing which may help their broader adoption in software testing automation. 

\end{abstract}


\maketitle

\section{Introduction}
Testing is one of the most critical phases of the software development life cycle (SDLC) \cite{10.1145/1764810.1764814,rastogi2015software,ruparelia2010software}. Software testing aims to detect requirements, design, and implementation faults so that a developer can fix them and demonstrate that the software performs desired behavior based on the specifications \cite{myers2011art,ammann2016introduction}. One key challenge in software testing is that it has become very costly and time-consuming due to the increasing complexity of modern software. Studies have found that approximately 50\% of the development time and more than 50\% of the total costs are spent on testing software \cite{myers2011art,10.1145/336512.336532}. Furthermore, testing is an incomplete process as it can \textit{not} show the absence of faults \cite{dahl1972structured}; therefore, given unlimited resources, one can run testing forever. However, in practice, only a limited resource is allocated to testing; thus, it must be stopped at some point. 

A \textit{test adequacy criterion} describes a set of rules that determines when sufficient testing has been performed and thus serves as a \textit{stopping rule}  for terminating a testing process \cite{10.1145/267580.267590}. Another purpose of test adequacy criteria is to assess the \textit{quality} of a test suite as, typically, a degree of adequacy is associated with each criterion \cite{10.1145/267580.267590}. Numerous test adequacy criteria, defined over program structures, specifications, or seeded faults \cite{10.1145/62959.62963,4527254,10.1145/267580.267590}, have been proposed, and researchers have investigated and evaluated the effectiveness of different criteria to support the use of one measure over another \cite{10.1145/62959.62963,296778}.

Code-based test adequacy criteria, also known as \textit{structural coverage criteria}, are defined over code structures such as statement, branch, condition, decision, function, def-use pair, etc.\cite{10.1145/62959.62963,hayhurst2001practical,296778,ntafos1988comparison}. For instance, the \textit{statement coverage} criterion is defined over program statements, requiring all statements to be executed at least once, and statement coverage is measured as the percentage of executed program statements\cite{10.1145/267580.267590}. Structural coverage criteria can be categorized into \textit{control-flow criteria} and \textit{data-flow criteria} \cite{10.1145/62959.62963,hayhurst2001practical}.

Data flow criteria are based on data flow (DF) analysis between the definitions and uses of program variables \cite{6194}. All-paths, all-du-paths, all-uses, all-p-uses, all-c-uses, all-p-uses/some-c-uses, all-edges, and all-nodes are some of the data-flow criteria proposed by Sandra Rapps, and Elaine J. Weyuker \cite{,1702019}. 
All-du-paths is one of the strongest data-flow criteria, which requires an adequate test suite to exercise \textit{every} path between the definition of a variable and each of its subsequent uses. In contrast, the all-uses criterion requires that \textit{at least} one path between the definition of a variable and its all uses be exercised under some test.


Control flow criteria are defined based on the control flow graph of a program. The basic control flow criterion is the statement criterion, which requires all control flow statements to be covered by a test suite. Even though statement coverage is the most used metric in practice, it is the weakest among all \cite{7272926}. Branch coverage criterion is stronger than the statement criterion as it requires all branches and all statements in the control flow graph to be covered \cite{7272926,zhu1997software}. Other stronger criteria are path coverage, decision coverage, condition/decision coverage, multiple condition/decision (MC/DC), and multiple condition coverage \cite{chilenski1994applicability,hayhurst2001practical,10.1145/267580.267590,7272926}. Multiple condition/decision (MC/DC) criterion is a requirement for testing safety-critical systems (RTCA DO-178) \cite{10.1145/3338906.3340459}.

Among test adequacy criteria, structural code coverage is popular in  industry because: (a) it is easy for developers to interpret; (b) support is available for most programming languages to integrate it into build processes; 
(c) it is efficient and introduces minimal overhead into test processes, and 
(d) many tools support its consumption by software engineers~\cite{berner2007enhancing,ivankovic2019code,8130777}. For these reasons, code coverage tools are used by millions of developers in thousands of organizations on a daily basis \cite{Codecov, LCOV}. 

Despite its strengths and popularity, code coverage has several limitations addressed by the research community \cite{schuler2013checked, zhang2015assertions}. Even though the sole purpose of testing is to detect faults, one major limitation of structural coverage criteria is that they do not meet the necessary conditions for a fault to be detected when it occurs while executing structural code elements. 
According to the PIE fault model ~\cite{voas1992pie}, for a fault to be detected, four conditions must be met: 
\begin{itemize}
    \item C1 -- Execution: Software structure containing fault must be executed.
    \item C2 -- Infection: The execution of the structure must result in an internal error state.
    \item C3 -- Propagation: The error state must propagate to program output.
    \item C4 -- Detection: The erroneous output must be observed and be judged to be incorrect.
\end{itemize} 


Traditional structural coverage criteria require only to satisfy the first condition, i.e., execution. However, executing a faulty program structure does not ensure that the fault is detected \cite{7272926,schuler2013checked}. To detect faults, one of the fundamental components of a test suite is \textit{test oracle}, a mechanism to check program output against expected output  ~\cite{staats2011programs,barr2014oracle}. One premise during the computation of code coverage is that test cases contain sufficient test oracles to detect faults. However, in practice, this assumption does not hold. Furthermore, as the computation of code coverage does not explicitly consider test oracles, it is entirely possible for a test suite with zero test oracles to achieve 100\% code coverage, resulting in a poor quality test suite with low fault-detection effectiveness.

A plethora of research verify the above theory that code coverage is inadequate as a test adequacy metric and is a poor indicator of a test suite fault-detection effectiveness. For example, through a large-scale study, Inozemtseva and Holmes \cite{inozemtsevaa2014notcorrelated,chen2020relationship} showed that the correlation
between code coverage and fault-detection effectiveness is weak. On the other hand, via large-scale studies, Zhang et al.  \cite{zhang2015assertions} showed that fault-detection effectiveness strongly correlates with test oracles. In summary, code coverage only measures the degree to which structural code elements are executed; it does not measure the extent to which test oracles check the results; therefore, using code coverage as a test adequacy metric overestimates the thoroughness of a test activity and test suite's fault-detection effectiveness.  

To address the limitations of code coverage, researchers proposed extensions to the traditional structural coverage that explicitly consider test oracles when computing code coverage. In general, in an oracle-based metric, an element from the coverage domain, e.g., statement, definition, MC/DC, is covered if it influences at least one value checked by a test oracle via a dependency chain. Therefore, an oracle-based metric ensures that the fault propagates to output and a test oracle checks the output against the expected output for fault detection. By doing so, oracle-based metrics satisfy C3 and C4 from the PIE model, increasing the likelihood of a fault being detected. However, a faulty program state can be masked out by subsequent program statements, and it is also possible that an oracle may not be strong enough to detect a fault.
Further oracle-based coverage ensures a mask-clear path from the executed code structure to test oracle. For example, \textit{observable coverage }\cite{ming2020observability,whalen2013observable} extends \texttt{boolean} expression-based coverage criteria, such as branch, condition, decision, condition/decision, and MC/DC, to ensure a mask-free path from the executed program structure to at least one oracle. However, for all oracle-based coverage, C4 from the PIE model is partially satisfied, as there is no guarantee that fault will be detected if it propagates to a test oracle. It is particularly challenging to distinguish, for a given input, the correct behavior from potentially an incorrect behavior, which is required for a test oracle to detect faults. This problem is a well-known problem in software testing, referred to as the ``test oracle problem''\cite{barr2014oracle}. Despite that, oracle-based coverage is shown to be way more effective than traditional coverage, as studies found a strong correlation between oracle-based coverage and fault-detection effectiveness \cite{zhang2015assertions,schuler2013checked}. 

Due to the significant importance of test oracles and oracle-based coverage in software testing, we have summarized all oracle-based coverage techniques published as of now. We have discussed the approaches thoroughly and presented their research findings. To the best of our knowledge, no other literature review study surveyed oracle-based coverage. Our paper provides valuable insight into the oracle-based coverage metrics that measure adequacy based on program execution and the program results checked by test oracles. 


\section{Mutation Coverage}
Mutation testing is a fault-based test adequacy metric that injects artificial faults in the program through minor modifications in the source codes to deviate from the original program's intended behavior, e.g., altering conditional boundaries, arithmetic operators, removing return values, etc. Mutation coverage domain is the set of generated faults (mutants). A mutant is killed/detected if any test fails when running on it, thus detecting the change. Mutation coverage is measured as the percentage of mutants detected by a test suite. A mutants is detected only when it is executed, an internal state is affected by the fault execution, the fault propagates to a test oracle and test oracle is strong enough to detect a fault. Therefore, mutation testing also fully satisfy C4 from the PIE model. 

Despite being a strong test adequacy metric, mutation testing has several limitations. One of the significant limitations is that it is costly to run mutation testing on a larger code base and test suite, as the for each generated mutant, the entire test suite is run on the source codes to detect a fault. Another limitation is equivalent mutants, which behave the same as the original program making it impossible to detect by any tests. Due to the nature of equivalent mutants, identifying them is an
NP-Complete problem.

\begin{table*}[t]
\caption{\label{tab-1} Different test adequacy metrics and their properties}
\begin{tabular}{|c|c|c|c|c|c|c|c}
\toprule
\textbf{Coverage Metric} & \thead{\textbf{Ensure} \\\textbf{Fault} \\\textbf{Execution}} & \thead{\textbf{Ensure} \\\textbf{Fault} \\\textbf{Propagation}} & \thead{\textbf{Ensure} \\\textbf{Fault} \\\textbf{Observation}} & \thead{\textbf{Ensure} \\\textbf{Fault} \\\textbf{Detection}} & \textbf{Coverage Domain} & \thead{\textbf{Underlying} \\\textbf{Technique }} \\

 \midrule
\textbf{Regular Coverage} & \checkmark  & \tikzxmark  & \tikzxmark & \tikzxmark& \thead{Program structure, \\e.g., statement, branch} & \thead{Program \\instrumentation} \\ \hline
\textbf{State Coverage \cite{koster2007state}} & \checkmark & \checkmark &  \thead{\checkmark \\(via dependency)} & \tikzxmark & \thead{Output defining \\and \\side effect variable} & \thead{Static program \\slicing/dynamic \\taint analysis} \\ \hline

\textbf{State Coverage \cite{vanoverberghe2013state}} & \checkmark & \checkmark & \thead{\checkmark \\(via dependency)} & \tikzxmark & \thead{State updates (writes),\\ object field only}& \thead{Dynamic \\information \\flow analysis} \\\hline

\textbf{Checked Coverage \cite{schuler2013checked}} & \checkmark & \checkmark & \thead{\checkmark \\(via dependency)} & \tikzxmark & \thead{Statements, \\Object Branch} & Dynamic slicing \\\hline

\thead{\textbf{Observable Coverage} \\ \textbf{\cite{whalen2013observable,ming2020observability}}} & \checkmark & \checkmark & \thead{\checkmark \\(mask-free)} & \tikzxmark & \thead{Boolean-expression, \\e,g., branch, \\condition, decision, \\MC/DC} & Tagged semantics\\\hline

\textbf{Mutation Coverage} & \checkmark & \checkmark & \checkmark & \checkmark & Seeded faults & 
\thead{Modification of \\program codes}\\

\bottomrule

\end{tabular}
\end{table*}

\section{Oracle-based code coverage}

The computation of oracle-based coverage (OBC) explicitly considers the presence and the quality of test oracles in a test suite. The quality is measures in terms of the percentage of coverage element that influence at least one value checked by test oracles.

\begin{definition}[\textbf{Oracle-based code coverage}]
For a particular coverage domain $H$ and a test suite $T$, $e(T)$ denotes the set of executed coverage elements from the coverage domain. OBC is computed as $|obc(T)|/|H|$, where $obc(.)$ only records elements from $e(.)$ on which at least one test oracle decision computation depends, i.e, there exists a transitive dependency between the output checked by oracle and the executed element in $e(.)$. $|H|$ denotes the size of the coverage domain. \end{definition}


Therefore, an element from the coverage domain is covered when it is executed and that code element affects a value checked by a test oracle via dependency chain (data and control dependency). The coverage element can be any code structure, such as statements, or boolean expression-based code structures, such as branch, condition, condition/decision.

For example, in checked coverage \cite{schuler2013checked}, a statement is covered if and only if that statement is executed and the statement influence at least one value checked by test oracles. Considering test oracles when computing code coverage has several benefits, such as if there is any fault in the covered statement, it is more likely that the statement will affect a test oracle and, therefore, more chances for that fault to be detected.

\section{Comparison of different test adequacy Metrics}
In Table \ref{tab-1}, we have shown properties of different test adequacy metrics, such as regular code coverage, oracle-based coverage, and fault-based mutation adequacy metrics. Traditional code coverage is the weakest test adequacy metric among all. Code coverage only satisfies the first necessary condition, i.e., execution,  for fault detection. Code coverage does not ensure that a fault (if it exists) is propagated to test oracles and detected by it. Code coverage domains are usually statement, branch, condition, decision, MC/DC, multiple conditions, def-use pairs, etc. Code coverage is typically computed by instrumenting the program under test and recording the executed program structures. 
Code coverage is usually computed by instrumenting the program under test and recording the executed program structures. 

Row 3-6 summarizes the properties of oracle-based adequacy metrics. All of them require to execute code elements from their respective domain and ensure that the executed coverage element influence/affect at least one value checked by a test oracle via data and control dependencies. Therefore, they all satisfy execution, infection (we assume that is true for all cases), propagation to output, and partially the fourth condition for fault detection. The fourth condition is partially satisfied as it is not guaranteed that fault will be detected just because the oracle is observing a value that depends on the executed coverage element. Fault detection also depends on the strength of the test oracle. Column six shows the respective coverage domain. The oracle-based metrics also vary in terms of their domain. For example, the state coverage domain is output-defining variables or state updates. Checked coverage and its extension considers the statement and object branch coverage as the coverage domain. On the other hand, observable coverage is interested in the Boolean-based structural elements, such as condition, decision, MC/DC, etc. For determining whether an executed code influence/affects any oracle decision, program slicing, taint analysis, or information flow analysis is used.  

Mutation coverage is the strongest among regular and oracle-based code coverage as a test adequacy metric. The coverage domain is seeded faults, artificially injected into the program via minor alteration of the program, e.g., altering conditional boundary, removing a return statement, etc. For a seeded fault to be covered, the fault must be executed, infect internal program states, propagate to output, and a test oracle must be strong enough to detect that fault. Due to this reason, mutation coverage is considered a strong test adequacy metric by several studies. One negative side of mutation testing is that it is costly and identifying equivalent mutants is an NP-Complete problem. Due to this reason, oracle-based coverage can be used as a viable option as it is more effective than regular coverage and less costly than mutation testing as the test suite needs to be executed only once, unlike mutation testing. Usually, in mutation testing, the program and the test suite must be executed $N$ times, where $N$ is the total injected faults. 

\section{LITERATURE REVIEW}
In this paper, we have discussed all publications on oracle-based coverage from 2007 to 2023. We found seven publications that proposed/extended oracle-based coverage when explicitly considering test oracles. We have discussed the following oracle-based coverage:

\begin{itemize}
    \item \textit{State coverage} by Ken Koster and David Kao, publish at ESEC/FSE’07 \cite{koster2007state}
    \item \textit{Checked coverage} by  David Schuler and Andreas Zeller, published at ICST'11 \cite{schuler2013checked}
    
    \item \textit{State coverage} by Vanoverberghe et al. \cite{vanoverberghe2011state}

    \item State coverage: An
empirical analysis based on a user study by Vanoverberghe et al., published at SOFSEM' 13 \cite{vanoverberghe2013state}
    
    \item \textit{Observable coverage}  by Whalen et al., published at ICSE'13 \cite{whalen2013observable}
    
    \item Extension of observable coverage -- Ensuring the Observability of Structural Test Obligations by Meng et al., published at TSE'17 \cite{8456606}

    \item Extension of checked coverage -- Measuring and Mitigating Gaps in Structural Testing by Hossain et al., published at ICSE'23
\cite{Hossain2023}

\end{itemize}


\subsection{\textit{State Coverage} by Koster et al. (FSE'07)}
In 2007, Ken Koster and David Kao proposed a structural test adequacy metric called \textit{state coverage}, published at ESEC/FSE’07. State coverage measures the percentage of output-defining variables and side effect variables checked by test oracles \cite{koster2007state}. Therefore, the coverage domain for state coverage is the set of output-defining variables and side effect variables. The output-defining variables define the output of a function/module, and side effect variables are usually non-local or static variables that are affected by the execution of some functions/methods. To the best of our knowledge, state coverage is the first paper that introduced a structural test adequacy metric that considers the use of test oracles in a test suite. By doing so, state coverage focuses more on checking the behavior of the programs than merely executing them. 

\subsubsection{\textbf{Methodology}}
An output-defining variable is determined based on the control flow graph (CFG) of the program under test (PUT) and the test. The PUT CFG is the subgraph of the total CFG. A CFG node consists of a statement, and the edges from that node indicate all possible control flow to another node. According to the definitions from the state coverage paper, $DEF(i)$ represents a set of variables defined at the CFG node $i$, $REF(i)$ is the set of variables used by the node at $i$. 

\begin{definition}[\textbf{Output-defining statement}]

Lets consider $n_t$ is the first node in the test ($t$) after executing PUT. If $n_d$ is a statement node from PUT that defines a variable $x\in DEF(n_d)$, and there is no redefinition of $x$ from $n_d$ to $n_t$, then $x$ is the output defining variable of <PUT,t> and $n_d$ is the output-defining statement.
\end{definition}

State coverage requires all output-defining variables from PUT concerning a test case $t$ to be checked by some test oracles. An output-defining variable $x$ is covered if test oracles check that variable or any other variables derived from it. To determine the CFG statement nodes and the variables checked by test oracles, \textit{program slicing} technique \cite{5010248} is used. Program slicing is a decomposition technique that extracts the relevant part of a program that may influence values at some point of interest, known as the \textit{slicing criterion}. A \textit{program slice} is a set of program statements that affect the slicing criterion. Program slicing can be computed for a particular input ( dynamic slicing) or all possible inputs (static slicing). State coverage uses the static slicing technique, and the variables being checked by the test oracles are used as the slicing criteria. An output-defining variable and the defining statement are covered only if at least one slice contains that statement.

The authors computed two types of state coverage: pessimistic and optimistic state coverage. Pessimistic state coverage required that all output defining statement of the PUT w.r.t test $t$ must be checked by test oracles in $t$. On the other hand, optimistic state coverage requires each output defining statement of the PUT be checked by at least one test in the entire test suite. 

\subsubsection{\textbf{Evaluation and Findings:} }
The authors conducted a small experiment on the \texttt{Apache Jakarta Commons Lang}, a utility package of Java. Commons Lang consists of 26 source classes with 1718 lines of code. The test suite consists of 309 test cases with a 74.5\% statement and condition coverage and 1607 test assertions. They randomly reduced the number of checks in a test suite to vary the total number of checks. They used \texttt{Indus}, a static program slicer for Java to compute static slices for the remaining checks in the test suite. Optimistic state coverage was calculated by dividing the total number of output-defining nodes in the slices by the total number of output-defining statements in the subject program. They also computed the mutation scores of different versions of the test suite using \texttt{muJava}, a mutation testing tool for Java.  They conducted a small study demonstrating the correlation between total checks, state coverage, and mutation score. The correlation between total checks and the mutation scores was much better than the correlation between checks and state coverage. Their results could be better if they used dynamic slicing instead of static slicing, as the dynamic slice is much smaller and more precise than the static slice.

\subsubsection{\textbf{State Coverage Tool (ICSE’08)}} In \cite{10.1145/1370175.1370210}, Ken Koster proposed a tool to compute state coverage for the Java JUnit test. Instead of using the static program slicing technique, they used dynamic taint analysis to compute state coverage. They considered all output defining statements as the source, and JUnit assertions as the critical sink for the taint analysis. The tainted sources that reach the assertions are considered covered, and others are considered uncovered. We were unable to find any open-source repository for the tool. 

\subsection{\textit{Checked Coverage} by Schuler et al. (ICST'11)}
In 2011, David Schuler and Andreas Zeller proposed \textit{checked coverage}, a metric to assess test oracle quality \cite{5770598,schuler2013checked}. Checked coverage is an extension to the statement coverage, and it measures the percentage of statements that are executed and that also influence the computation of test oracles. When there are only a few test oracles or poor quality oracles that check only a little, checked coverage drops quickly; thus, it indicates the quality of test oracles in the test suite.  

\subsubsection{\textbf{Methodology}} The computation of checked coverage requires to find the statements that are needed to compute the results checked by test oracles. To this end, they used \textit{dynamic program slicing}. This technique extracts the data and control-dependent statements that are required to compute the value of a variable at a program point (known as the slicing criterion) for a particular program input. They considered all test oracles in a test suite as the slicing criteria and constructed dynamic program slices consisting of all statements required to compute the values checked by those test oracles. Then checked coverage is calculated as the ratio of the total statements in the slices to the total executed statements by the test suite. 

To implement checked coverage, JavaSlicer \cite{hammacher2008design}, a backward dynamic program slicing tool for Java is used. JavaSlicer computes slices in two steps: tracing and slicing. First, JavaSlicer instruments all java classes to collect the execution traces by running the JUnit test suite. Second, for a specific test oracles, i.e., slicing criterion, it computes the dynamic slice by traversing the execution trace and transitively collecting the data and control dependencies. The resulting slice is a set of statements required to compute the value of the variable checked by the test oracle. Checked coverage for an entire test suite is computed by forming slicing criteria for all test oracles in the test suite, computing dynamic slices for those criteria, and then dividing the total number of unique statements in the slices by the total number of executed statements provides the checked coverage value.  

\subsubsection{\textbf{Evaluation and Findings:}} Checked coverage was evaluated on seven open-source Java applications. Their experimental results showed that checked coverage is always lower than statement coverage, and there is an average difference of 24\%, meaning that 24\% of the executed statements do not influence any test oracle, i.e., no test oracles check them. 

They also conducted another study to compare the sensitivity of different adequacy metrics, such as checked coverage, statement coverage, and mutation score, to removed test oracles. To this end, they systematically removed 0\%, 25\%,50\%,75\%, and 100\% test assertions from the test suites and computed coverage metrics for different versions of test suites. They found that all metrics, checked coverage, statement coverage, and mutation coverage drops when assertions are removed from test suites. However, when a test assertion is removed, statement coverage drop depends on the presence of method calls inside the assertion body. Removing an assertion does not affect the statement coverage when there is no such method call. However, it may affect the checked and mutation coverage. Their experimental results showed that checked coverage is more sensitive to assertion removal than statement and mutation coverage.


\subsection{\textbf{State Coverage by Vanoverberghe et al. (SOFSEM'12)} }
In 2012, Vanoverberghe et al. extended the original definition of the state coverage proposed by Ken Koster and David Kao \cite{koster2007state}. In contrast to the original definition of state coverage \cite{koster2007state}, which requires checking all output-defining variables by test assertions, state coverage requires all state updates (writes) to be checked (read) by test assertions \cite{vanoverberghe2011state}. This metric is most similar to the all-defs \cite{rapps1985selecting} coverage criterion, which requires every definition to be used (read) by at least one statement node \cite{10.1145/62959.62963}. Whereas the state coverage requires all definitions (states) to be read by at least one assertion.

\begin{definition}[\textbf{State coverage}]
State coverage measures the percentage of program states read by test assertion w.r.t to the total number of state updates.
\end{definition}

State coverage can be computed in two granularities: object insensitive and sensitive state coverage. 

\begin{definition}[\textbf{Object insensitive state coverage}]
Object insensitive state coverage considers state updates irrespective of the object on which the update was performed (similar to statement coverage) and measures the percentage of state updates checked by assertions.
\end{definition}

\begin{definition}[\textbf{Object sensitive state coverage}]
Object-sensitive state coverage is sensitive to object whose state is being updated. A state update is a pair of code location that performs an update and the object identifier whose state has been updated 
\end{definition}

\subsubsection{\textbf{Methodology} }
For computing object-insensitive state coverage, a runtime monitor is implemented, which keeps track of the object fields that are written and the object fields that are read by test assertions. When a field is written, a set \textit{writes} stores the code location of that write, and a map \textit{lastWrite} updates the last location of the write for that field. Note that a code location is a pair of a method and the offset of an instruction inside the method body where the write operation has been performed. When program execution invokes any assert method, all fields that are read from the assert body are recorded, and the last location of those fields are added to another set \textit{reads}.

Object-sensitive state coverage requires more information than object-insensitive coverage because the read and write to fields may happen to the same code location but on different objects. To this end, they assigned a context-insensitive object identifier to each object to keep track of the writes and reads to the fields of a particular object and code location. 
Resulting \textit{write} and \textit{read} sets over multiple tests can be unioned to report a composable object-sensitive state coverage for an entire test suite. 

They implemented state coverage as an extension to Pex \cite{tillmann2008pex}, a dynamic symbolic execution-based input generation tool for the .NET framework used at Microsoft Research. Even though Pex is publicly available as a Visual Studio plugin, we are not sure whether the implementation of state coverage is also publicly available.

\subsubsection{\textbf{Evaluation and Findings:}} State coverage was evaluated on two open-source libraries: \texttt{Quickgraph} and  \texttt{Data Structures and Algorithms (DSA)}. The authors did not conduct any quantitative analysis; however, they conducted a small qualitative study to investigate whether increasing state coverage for programs with high structural coverage impacts the number of bugs found. To this end, they manually added assertions to improve state coverage by reading fields that were written but not read by any assertions. They reported the total number of assertions added and bugs found due to additional assertions when the state coverage improvement reached saturation. For the DSA project, they added 33 additional assertions and found five new bugs, whereas for the Quickgraph, they added 22 assertions and found no additional bugs. Finally, for the DSA subject, they thoroughly discussed the type of pre and post-conditions and invariants they enforced to write the assertions, which helped detect bugs in the BinarySearchTree and DoublyLinkedList from the DSA project.

In 2013, Vanoverberghe et al. published an empirical study on the evaluation of the state coverage \cite{vanoverberghe2013state}. The author conducted a user-based experiment to assess the relationship between state coverage and a test suite fault-detection effectiveness. To this end, they manually injected faults and asked external users to write test cases to detect them. Next, they computed state coverage and the number of faults detected. Their experimental results failed to confirm any significant correlation between these two variables. The authors mentioned that state coverage is most suitable for finding logical faults that occur on rare occasions, which is one reason the hypothesis failed. Furthermore, adding new test cases can decrease state coverage even though new properties are checked and bugs are found. Due to this reason, any positive correlation between state coverage and fault detection could not be confirmed.

\subsection{Observable Coverage by Whalen et al. (ICSE'13)}

Among all structural coverage criteria, the modified condition/decision coverage (MC/DC) criterion is considered one of the strongest criteria required by the safety-critical domains (RTCA DO-178) \cite{10.1145/3338906.3340459,hayhurst2001practical}. However, like any other structural test adequacy metric, MC/DC has several limitations. For example, satisfying MC/DC obligation only requires each condition within a decision to be exercised in both directions (true and false) and independently affect the decision value \cite{hayhurst2001practical}. It does not guarantee that faults will be propagated and detected by test oracles. Another limitation is that in non-inlined implementation, corrupted internal states can be masked out by subsequent conditions, preventing the faults to be propagated to the output. Due to these reasons, the effectiveness of the MC/DC metric depends on program structures (inline/non-inline), which can vary to a high degree across different groups. 

In 2013, Whalen et al. proposed \textit{observability}, an extension to the regular modified condition/decision coverage (MC/DC) criterion \cite{whalen2013observable}. Observable MC/DC (OMC/DC) adds an extra condition to the MC/DC requirement to ensure a non-masking path from the executed program condition to a test oracle that monitors the output. Even though the additional conditional makes the MC/DC requirements difficult to satisfy, it increases the likelihood of fault propagation to a test oracle for better fault detection effectiveness. 

\subsubsection{\textbf{Methodology}}

The implementation of observable MC/DC used tagging semantics -- an approximation-based forward analysis technique to determine the observability. Each atomic condition is marked with a tag, where a tag is a pair of values: unique id and the boolean value of the condition. After running the test suite, if a tag associated with a condition reaches at least one of the variables monitored by test oracles, the test satisfies the observability condition. Observable MC/DC is computed by dividing the total tags reached to output by the total number of tags in the program. 

\subsubsection{\textbf{Evaluation and Findings} }
To evaluate OMC/DC, the authors conducted experimental studies on four systems from Rockwell Collins. For each system, they generated inlined and non-inlined versions and 10 test suites satisfying OMC/DC and MC/DC. Mutation testing was used to assess the fault-detection effectiveness of different test suites. Their experimental results found that OMC/DC outperformed MC/DC and detected a median of 17.5\% and up to 88\% more faults than test suites providing MC/DC coverage when paired with output oracles. Furthermore, their study also demonstrated that OMC/DC is less sensitive to program structure. When converting a program structure from non-inline to inline, MC/DC effectiveness for output-only oracles increases significantly. However, OMC/DC improvements are much lower, indicating that OMC/DC is less sensitive to program structures. When comparing the test suite size and test case length, OMC/DC requires more tests for non-inlined programs, and the test case lengths are also longer than MC/DC.

\subsection{\textbf{Observable Structural Coverage}}
In 2017, Meng et al. \cite{8456606} extended the concept of observability for other \texttt{Boolean} expression-based structural coverage criteria, such as Branch, Condition, Decision, and MC/DC. Similar to the original definition of observability defined in \cite{whalen2013observable}, an additional condition is imposed with the regular host coverage to ensure a masking-free path from the executed structural element to a test oracle that monitors the output. This increases the likelihood of propagating a fault to test oracles and be detected, thus eliminating the limitation of regular structural coverage that only ensures that a particular program structure is executed. 

Similar to \cite{whalen2013observable}, they used tagging semantics-based forward analysis technique to determine whether a tag for an atomic condition is observable to a test oracle. If some other conditions mask the Boolean value of the condition under test, then it will not be observable in the output variable monitored by test oracles. 

\subsubsection{\textbf{Evaluation and Findings}} To evaluate observable coverage, two sets of programs written in Lustre language were used. The authors used mutation testing and generated 500 mutants with a single fault to assess and compare the fault-detection effectiveness of the regular structural coverage and their observable variants. They generated test suites for each coverage criterion and 50 reduced test suites from the main test suites and computed the fault-detection effectiveness via mutation testing. Their experimental results showed that observable coverage is way more effective than regular coverage, and OMC/DC is most effective among all observable and non-observable variants. For maximum oracle (checks all internal variables), adding observability improves fault detection by 11.94\% on average, over regular coverage and 125.98\% for output-only oracles. 

Satisfying observability requires more extended tests and results in increased size in the test suite. The magnitude of the increase in the test suite size depends on the length of the path from each boolean expression to the output. Furthermore, the addition of observability also results in a decrease in the number of fulfilled obligations due to the complexity of the test obligations or dead code that can not influence any outputs.

\subsection{\textbf{Extension of Checked Coverage}}
At ICSE'23 \cite{Hossain2023}, Hossain et al. proposed an extension and improvement over checked coverage (CC) by David Schuler and Andreas Zeller \cite{schuler2013checked}. CC only supports statement criterion and measures the percentage of executed program statements that influence at least one value checked by test oracles. As the coverage domain of CC and the statement coverage is different, comparing these metrics is not straightforward. To address this, Hossain et al. redefined checked coverage for the statement criterion and proposed an extension to the checked coverage for the object branch criterion. Their extension to CC is known as HCC, host checked coverage. 

\begin{definition} [\textbf{Host Checked Coverage (HCC)}]
For a particular coverage domain $H$ and a test suite $T$, $e(T)$ denotes the set of executed coverage elements. HCC is computed as $|hcc(T)|/|H|$, where $hcc(.)$ only records elements from $e(.)$ that influence at least one test oracle decision in $T$. $|H|$ denotes the size of the coverage domain. For example, SC denotes statement coverage, and SCC denotes statement checked coverage. Similarly,  OBC represents object branch coverage, and OBCC represents object branch checked coverage.
\end{definition}

\begin{definition} [\textbf{Coverage Gap ($G_H$)}]
Since $HC$ and $HCC$ are defined over the same domain, $H$, gap between these two metric can be computed as the difference between these two values and represent as percentage points (PP). For a test suite $T$, coverage domain $H$, coverage gap 
$G_H(T) = H(T) \setminus HCC(T)$
\end{definition}

Like CC, HCC also used JavaSlicer, a dynamic backward slicing tool for Java programs to determine which elements from the coverage domain are executed and influencing at least one value checked by test oracles. All test oracles in the test suite are considered as slicing criteria to compute program slices which consist of all data and control dependencies to compute the values in the slicing criteria. HCC is then calculated as the percentage of coverage elements in the slices w.r.t the total coverage elements in the program. For example, statement checked coverage is the ratio of the total statements in the slice to the total statements in the program.

\subsubsection{\textbf{Evaluation and Findings}} 
HCC is evaluated on 13 large-scale Java artifacts. 
Their study revealed that even for mature and well-developed test suites, there is a considerable gap between HCC and HC -- traditional structural coverage for statement and object branch criteria. They investigated the correlation between the coverage gap and the fault-detection effectiveness of test suites. To this end, they generated several version test suites with varying coverage gaps and performed mutation testing to measure their fault-detection effectiveness. 
Their experimental results showed a strong negative correlation between the coverage gap and the fault-detection effectiveness of test suites for both statement and object branch criteria in multiple granularities (whole application, package).


\section{Conclusion}
In this survey paper, we have discussed a total of seven publications on the oracle-based structural test adequacy metrics, published from 2007 to 2023. To the best of our knowledge this is the first survey paper in this area. Oracle-based coverage metrics have shown to be effective in detecting more faults than regular coverage metrics, however, they have received only little attention among the software engineering research and practitioner community. Our servery paper has put all information together for easy access and understanding.  

\bibliographystyle{plainnat}
\bibliography{survey}
\end{document}